\documentclass[aps,twocolumn,superscriptaddress,notitlepage,nofootinbib]{revtex4-1}

\usepackage{amsmath,bm}
\usepackage{graphicx}
\usepackage{epstopdf}
\usepackage{subfigure}
\usepackage{epsfig}
\usepackage{amsmath,amssymb,amsfonts}
\usepackage{color}
\usepackage[utf8]{inputenc}
\usepackage{verbatim}
\usepackage{array}
\usepackage{hyperref}
\usepackage{ulem}
\usepackage{multirow}

\begin{document}

\renewcommand{\topfraction}{0.85}
\renewcommand{\textfraction}{0.1}
\renewcommand{\floatpagefraction}{0.75}

\title{Flavor dependence of jet quenching in heavy-ion collisions from a Bayesian analysis}

\date{\today  \hspace{1ex}}

\author{Shan-Liang Zhang}
\email{Corresponding author.\\
zhangshanl@m.scnu.edu.cn}
\affiliation{Key Laboratory of Atomic and Subatomic Structure and Quantum Control (MOE), Guangdong Basic Research Center of Excellence for Structure and Fundamental Interactions of Matter, Institute of Quantum Matter, South China Normal University, Guangzhou 510006, China}
\affiliation{Guangdong-Hong Kong Joint Laboratory of Quantum Matter, Guangdong Provincial Key Laboratory of Nuclear Science, Southern Nuclear Science Computing Center, South China Normal University, Guangzhou 510006, China}
\affiliation{
Department of Physics, Hubei University, Wuhan 430062, China
}

\author{Enke Wang}
\email{wangek@scnu.edu.cn}
\affiliation{Key Laboratory of Atomic and Subatomic Structure and Quantum Control (MOE), Guangdong Basic Research Center of Excellence for Structure and Fundamental Interactions of Matter, Institute of Quantum Matter, South China Normal University, Guangzhou 510006, China}
\affiliation{Guangdong-Hong Kong Joint Laboratory of Quantum Matter, Guangdong Provincial Key Laboratory of Nuclear Science, Southern Nuclear Science Computing Center, South China Normal University, Guangzhou 510006, China}

\author{Hongxi Xing}
\email{Corresponding author.\\
hxing@m.scnu.edu.cn}
\affiliation{Key Laboratory of Atomic and Subatomic Structure and Quantum Control (MOE), Guangdong Basic Research Center of Excellence for Structure and Fundamental Interactions of Matter, Institute of Quantum Matter, South China Normal University, Guangzhou 510006, China}
\affiliation{Guangdong-Hong Kong Joint Laboratory of Quantum Matter, Guangdong Provincial Key Laboratory of Nuclear Science, Southern Nuclear Science Computing Center, South China Normal University, Guangzhou 510006, China}
\affiliation{Southern Center for Nuclear-Science Theory (SCNT), Institute of Modern Physics, Chinese Academy of Sciences, Huizhou 516000, China}

\author{Ben-Wei Zhang}\email[]{bwzhang@mail.ccnu.edu.cn}
\affiliation{Institute of Particle Physics and Key Laboratory of Quarks and Lepton Physics (MOE), Central China Normal University, Wuhan 430079, China}

\begin{abstract}
We investigate the flavor dependence of jet quenching, by performing a systematic analysis of medium modifications on the inclusive jet, $\gamma$+jet, and $b$-jet in Pb+Pb collisions at the LHC. Our results from MadGraph+PYTHIA exhibit excellent agreement with experimental measurements of the inclusive jet, $\gamma$+jet and $b$-jet simultaneously  in p+p collisions. We then utilize a Bayesian data-driven method to extract systematically the flavor-dependent jet energy loss distributions from experimental data, where the gluon, light quark and $b$-quark initiated energy loss distributions are well constrained and satisfy the predicted flavor hierarchy of jet quenching, i.e. $\langle \Delta E_g \rangle > \langle\Delta E_q\rangle > \langle\Delta E_b\rangle$. It is shown that the 
quark-initiated jet energy loss distribution shows weaker centrality and $p_\text{T}$ dependence than the gluon-initiated one. We demonstrate the impacts of the slope of initial spectra,
color-charge as well as parton mass dependent jet energy attenuation on the $\gamma/b$-jet suppression observed in heavy-ion collisions.
  %We find that large quark-initiated jet fraction contribute $80\%$ to large $p_\text{T}$ $\gamma$+jet suppression, while the flat spectra have the dominate contributions to low $p_\text{T}$ $\gamma$+jet suppression.
  %Furthermore, $b$-initiated jets is less suppressed compared to light-quark initiated jets,  but the mass effects  decrease as $p_\text{T}$.
   %Except for mass effect, smaller fraction of gluon contributions to $b$-jet also lead to less suppression of $b$-jet compared to inclusive jet in low $p_\text{T}$ jet region.

\end{abstract}


\maketitle

%\section{introduction}
%\label{sec:Intro}

\section{Introduction}
%Jet quenching has long been proposed as an excellent probe of the properties of the Quark-Gluon Plasma (QGP) \cite{Wang:1991xy,Wang:1998ww}, formed in high-energy heavy-ion collisions at Relativistic Heavy Ion Collider (RHIC)\cite{Adcox:2001jp,Adler:2002xw, Adler:2002tq} and the Large Hadron Collider (LHC) \cite{Aad:2010bu,Chatrchyan:2012gt,Chatrchyan:2013kwa,Aad:2014bxa}.
The understanding of strongly interacting nuclear matter at extremely high temperature and energy density is one of the primary subjects in the study of high-energy nuclear collisions at Relativistic Heavy Ion Collider (RHIC)\cite{Adler:2002xw, Adler:2002tq} and the Large Hadron Collider (LHC) \cite{Aad:2010bu,Chatrchyan:2012gt,Chatrchyan:2013kwa,Aad:2014bxa}. Jet quenching has long been identified as a very powerful tool to investigate the phase transition from hadron gas to  the quark-gluon plasma (QGP) with deconfined quarks and gluons~\cite{ Wang:1991xy, Wang:1998ww}, and numerous studies have shown that parton energy loss in the QGP  may lead to the  suppression of the single inclusive hadron/jet spectra~\cite{Qin:2007rn,Chen:2011vt,Buzzatti:2011vt,Majumder:2011uk,Aamodt:2010jd,CMS:2012aa,Qiu:2019sfj},
 the shift of $\gamma$/Z+hadron/jet correlations \cite{Wang:1996yh,Renk:2006qg,Zhang:2009rn,Qin:2009bk,Adare:2009vd,Abelev:2009gu,Chen:2017zte,Dai:2012am,Neufeld:2010fj,Kang:2017xnc,Luo:2018pto,Zhang:2018urd,Sirunyan:2017jic,Yang:2021qtl,Zhang:2021oki,Zhang:2022bhq} and dihadron transverse momentum asymmetry~\cite{Zhang:2007ja,stardihadron,Ayala:2009fe,Ayala:2011ii}, the modification of jet internal structures~\cite{ATLAS:2019dsv,Zhang:2021sua,Chang:2019sae,Chang:2019sae,KunnawalkamElayavalli:2017hxo,Kang:2017mda,Ma:2013uqa,Chatrchyan:2014ava,Aaboud:2017bzv,Vitev:2008rz}, as well as the azimuthal anisotropy ($v_2$) of hadrons and jets~\cite{STAR:2003wqp,CMS:2017xgk,ATLAS:2018ezv,He:2022evt} with the large transverse momentum ($p_\text{T}$) in nucleus-nucleus (A+A) collisions, by comparison with those in proton-proton (p+p) collisions~\cite{Qin:2015srf,CMS:2021vui,Apolinario:2022vzg}.

The interaction between an energetic parton and the QGP is sensitive to the colour charge and the mass of the parton, while medium-induced gluon
radiation is expected to be enhanced for gluon due to its larger color factor, and  to be suppressed for heavy quarks by the dead-cone effect relative to that for 
light quarks~\cite{Dokshitzer:2001zm,Zhang:2003wk,Djordjevic:2003qk,Armesto:2003jh}.
%While comprehensive efforts have been devoted to extract key transport properties of QGP based on jet quenching frameworks (see e.g. by JET and JETSCAPE collaborations \cite{JET:2013cls,JETSCAPE:2021ehl,JETSCAPE:2020shq}), the information on the jet quenching of specific parton type is still limited.
Such a predicted flavor hierachy of jet quenching can be identified by a separate determination of quark and gluon jet energy loss, which could play a significant role in revealing the fundamental color structures of the QGP and testing the color representation dependence of the jet-medium interaction~\cite{Frye:2017yrw,Gras:2017jty}. This however proves difficult, as the final state hadronic observables are a mixture of quark and gluon contributions. A clean method for identifying quark or gluon energy loss remains a challenge, despite many past attempts such as the multivariate analysis of jet substructure observables~\cite{Chien:2018dfn}, the proposal of using the averaged jet charge~\cite{Chen:2019gqo,CMS:2020plq,Li:2019dre}  and electroweak gauge boson tagged jet~\cite{Dai:2012am,Neufeld:2010fj,Kang:2017xnc,Luo:2018pto,Zhang:2018urd,CMS:2017eqd,Yang:2021qtl, Yan:2020zrz,Zhang:2021oki,Aaboud:2019oac}.

One recent important measurement by the ATLAS Collaboration, i.e. the  nuclear modification factor for  $\gamma$-tagged and $b$-tagged jets~\cite{ATLAS:2022cim,ATLAS:2022fgb}, shows quite different modification pattern from that of single inclusive jets~\cite{ATLAS:2018gwx}. It is reported that the $\gamma$-tagged jets $R_\text{AA}$~\cite{ATLAS:2022cim} are much higher and show a weaker centrality dependence than inclusive jet $R_\text{AA}$~\cite{ATLAS:2018gwx} , indicating a sensitive observation of color factor dependence of jet-medium interaction. In addition, the ratio of $R_\text{AA}$ between $\gamma$-tagged jet and inclusive jets are above most of the theoretical model calculations~\cite{ATLAS:2022cim}, which challenges the implemented color charge dependence of energy loss in these models. Likewise, systematic difference between $b$-jets and inclusive jets $R_\text{AA}$ are also observed~\cite{ATLAS:2022fgb} and , suggesting a role for mass and colour charge effects in partonic energy loss in heavy-ion collisions. Those differences may arise from not only the inclusive jet mixture of quarks and gluons, where gluon lose more energy, but also the slope of their initial spectrum~\cite{He:2018xjv}. 
 Meanwhile, most theoretical models can capture the inclusive jet $R_\text{AA}$~\cite{ATLAS:2018gwx}. However,  discrepancies arise when examining the latest photon/b-tagged jet $R_\text{AA}$ data points and the double ratio $R^{\gamma/b\text{+jet}}_\text{AA}/R^\text{inclusive jet}_\text{AA}$~\cite{ATLAS:2022cim,ATLAS:2022fgb}. In general, these quantities tend to surpass the predictions of many jet quenching models grounded in pQCD calculations and kinetic theory. Such contradictions strongly motivate a data-driven Bayesian analysis to extract the model-independent yet flavor-dependent jet energy loss distributions, which can not only identify the transport properties of QGP~\cite{Zhang:2022rby}, but also help to pin down the uncertainties of jet quenching models.
%\sout{Therefore, it is necessary to have explicit knowledge of model-independent but flavor-dependent jet energy loss distributions, which can help to constrain jet quenching model uncertainties and to identify the transport properties of QGP~\cite{Zhang:2022rby}.}

The purpose of this work is to extract the flavor-dependent and  model-independent jet energy loss distributions by performing a systematic study of the medium suppression of  the inclusive jet, $\gamma$+jet, and $b$-jet in Pb+Pb collisions relative to that in p+p in a unified framework simultaneously. In the numerical calculation of the p+p baseline, a Monte Carlo event generator MadGraph5+PYTHIA8~\cite{Alwall:2014hca}, which can perform the next-to-leading order (NLO) matrix element (ME) matched to the resummation of parton shower (PS) calculations, is employed to simulate initial hard partons with shower partons and jet cross sections in p+p collisions.
Specifically, a Bayesian data-driven analysis~\cite{He:2018gks} of the nuclear modification factors of inclusive jet~\cite{ATLAS:2018gwx}, $\gamma$+jet~\cite{ATLAS:2022cim}, and $b$-jet~\cite{ATLAS:2022fgb} is performed to quantitatively extract the flavor dependent jet energy loss distributions, which satisfies the predicted flavor hierarchy of jet quenching. We study the relative contributions from the slope of initial spectra, color-charge as well as parton mass dependent jet energy attenuation to the $\gamma/b$-jet suppression in heavy-ion collisions at the same time.

The remainder of the paper is organized as follows.  In Sec.~\ref{sec:Fra} we first introduce the framework.
%And then we present the evaluation of the inclusive jet, $\gamma$+jet, and $b$-jet cross section in p+p collisions using MadGarph+Pythia~\cite{Alwall:2014hca} simulations
With a systematic study of the inclusive jet,  $\gamma$+jet, and  $b$-jet productions in p+p collisions using MadGarph+Pythia, a Bayesian data-driven analysis of nuclear modification factors of these processes is performed to quantitatively extract flavor dependent jet energy loss distributions in Sec.~\ref{sec:nume}.  Finally, a summary is presented in Sec.\ref{summary}.

%\label{numerical}
%\section{Results from LBT simulation}
\section{Framework}
\label{sec:Fra}
In order to study the flavor dependence of jet energy loss, we express the final observable of the nuclear modification factor $R_\text{AA}$ in a given centrality in terms of the flavor dependent $R_\text{AA}^{i,C}$,
\begin{equation}
\begin{split}\label{EQ_raa}
  R_\text{AA}^{C}&= \frac{\sum_i R_\text{AA}^{i,C}d\sigma^{i}_\text{pp} }{\sum_id\sigma^{i}_\text{pp}}=R_\text{AA}^{g,C}+ \sum_{i\neq g}(R_\text{AA}^{i,C}-R_\text{AA}^{g,C} )f_{i},
 % &= \frac{\sigma_{c\rightarrow J/\psi}R_\text{AA}^c+\sigma_{g\rightarrow J/\psi}R_\text{AA}^g}{\sigma_{c\rightarrow J/\psi}+\sigma_{g\rightarrow J/\psi} }\\
 % &=R_\text{AA}^g+\frac{\sigma_{c\rightarrow J/\psi}(R_\text{AA}^c-R_\text{AA}^g )}{\sigma_{c\rightarrow J/\psi}+\sigma_{g\rightarrow J/\psi}}\\
 % &=R_\text{AA}^g+ (R_\text{AA}^q-R_\text{AA}^g )f_{q}
\end{split}
\end{equation}
%where $\sigma^{i}_\text{pp}$ is the cross section for parton $i$ in $p+p$ collisions, and $\sum_i \sigma^{i}_\text{pp}$ corresponds to the final observed cross section in $p+p$ collisions.
where the superscripts $i$ and $C$ stand for the parton flavor and centrality, respectively. $d\sigma^{i}_\text{pp}$ is the differential cross section for parton $i$ initiated jet in p+p collisions, $f_{i}=d\sigma^{i}_\text{pp}/\sum_id\sigma_\text{pp}^i$ is the fraction of the total jet cross section from the parton $i$ initiated one.

In our analysis, the flavor and centrality dependent nuclear modification factor $R_\text{AA}^{i,C}$ is assumed to be factorized as the convolution of its cross section in p+p collisions and the corresponding parton energy loss distribution~\cite{He:2018xjv,He:2018gks}
%\begin{equation}\label{straight}
%\frac{d \sigma^{g^\prime}}{ dp^{g^\prime}_\text{T}}=   \int \frac{dp_\text{T}}{\langle \Delta p_\text{T}\rangle} \frac{d \sigma^{pp}}{dp_\text{T}}(p_\text{T} )  \times W_\text{AA}(x)
%\end{equation}
\begin{equation}
R_\text{AA}^{i,C}(p_\text{T})= \frac{\int d\Delta p_\text{T} d\sigma_\text{pp}^i(p_\text{T}+\Delta p_\text{T})\otimes W_\text{AA}^{i,C}(x)}{d\sigma_\text{pp}^i(p_\text{T})},
\label{sigma_AQ}
\end{equation}
where $x=\Delta p_\text{T}/\langle \Delta p_\text{T}\rangle$ is the scaled variable with $\Delta p_\text{T}$ the amount of energy loss and $\langle \Delta p_\text{T}\rangle$ the averaged jet energy loss, which can be parametrized as $\langle \Delta p_\text{T}\rangle=\beta_i (p_\text{T})^{\gamma_i} \log(p_\text{T})$ following Refs.~\cite{He:2018xjv,Zhang:2021sua}.
 %We assume that the centrality and $x$ dependence of the energy loss distribution can be factorized as $W_\text{AA}^{i,C}(x)=\omega^{i,C}W^i_\text{AA}(x)$, where $f^{i,C}$ is the centrality dependent part, while $W^i_\text{AA}(x)=\frac{\alpha_i^{\alpha_i} x^{\alpha_i-1}e^{-\alpha_i x} }{\Gamma(\alpha_i)}$ is the scaled centrality-independent energy loss distributions of parton $i$.
In Eq. (\ref{sigma_AQ}), $W_\text{AA}^{i,C}$  is the scaled energy loss distribution of parton $i$ in a given centrality class $C$ of A+A collisions and can be assumed as:
\begin{equation}
W^{i}_\text{AA}(x)=\frac{\alpha_{i}^{\alpha_{i}} x^{\alpha_{i}-1}e^{-\alpha_{i} x} }{\Gamma(\alpha_{i})}
\label{Waa}
\end{equation}
where $\Gamma$ is the standard Gamma-function, and the above functional form can be empirically interpreted as the energy loss distribution
resulting from $\alpha_i$ number of jet-medium scattering in the medium.

In this setup, for each parton flavor $i$, the scaled jet energy loss distributions  $W_\text{AA}^{i}(x)$  can be determined by three parameters,  $\alpha_i, \beta_i, \gamma_i$. According to this flavor decomposition, one can extract $\alpha_i, \beta_i, \gamma_i,$ for each parton flavor $i$ to determine the flavor and  centrality dependent jet energy loss distributions  $W_\text{AA}^{i}(x)$ through a global analysis by combining the simulations of p+p cross section and the measurements of nuclear modification factor $R_\text{AA}$ for jet related observables.

We apply an advanced statistical tool, i.e. Bayesian analysis, for this purpose. Such a method has been successfully employed to extract the bulk and heavy quark~\cite{Xu:2017obm},  jet~\cite{He:2018gks} and gluon~\cite{Zhang:2022rby} energy loss distributions as well as  jet transport coefficients~\cite{Xie:2022ght,JETSCAPE:2021ehl} in heavy-ion collisions. The process can be summarized as
\begin{equation}
P(\theta|data) =\frac{P(\theta)P(data|\theta) }{ P(data)},
\label{Bayesian}
\end{equation}
where $P(\theta|data)$ is the posterior distribution of parameters $\theta$ given the experimental data, $P(\theta)$
is the prior distribution of $\theta$, $P(data|\theta)$ is the Gaussian likelihood between experimental data and the output for any given set of parameters and $P(data)$ is the evidence.  Uncorrelated uncertainties in experimental data are used in the evaluation
of the Gaussian likelihood.  To estimate the posterior distribution given by Eq.~\ref{Bayesian},
the Markov chain Monte Carlo (MCMC) process is carried out using the Methropolis-Hasting algorithm~\cite{Andrieu}. A uniform prior distribution $P(\theta)$ in the region $[\alpha_i, \beta_i, \gamma_i] \in [(0,10),(0,8),(0,0.8)]$ is used for the  Bayesian analysis.  We
first run $2 \times 10^6$ burn-in MCMC steps to allow the chain
to reach equilibrium, and then generate $2 \times 10^6$ MCMC steps
in parameter space. 
%{\color{red}It may be noted that the Bayesian analysis here uses specific functional form for the parameterzation, thus introducing long-range correlations in the parameter space which may potentially bias the extracted parameters. A possible solution to tackle such issue is to use information field based approach as presented in Ref.~\cite{Xie:2022ght}.}
%\textcolor{red}{what is $P(data)$? what's the prior distribution of $\omega_i^C$?}

\section{Results and Discussions}
\label{sec:nume}

\subsection{Cross sections in p+p }
In our analysis, we consider three different observables, i.e. the inclusive jet, $\gamma$+jet and $b$-jet, to study the flavor dependence of jet energy loss distribution. Considering the facts NLO matching have considerable contributions to  b-jet cross section~\cite{Banfi:2007gu}  and $\gamma$+jet~\cite{Zhang:2018urd},  we simulate $d\sigma^{i}_\text{pp}$ using a Monte Carlo event generator MadGraph5+PYTHIA8~\cite{Alwall:2014hca}, which combines the NLO matrix element (ME) with the matched parton shower (PS). Furthermore, those shower partons are reconstructed using the anti-$k_\text{T}$  algorithm~\cite{Cacciari:2008gp} implemented in the FastJet~\cite{Cacciari:2011ma}. In order to compare with the $b$-jet measurements, we define $b$-jet to be the one that contains at least one $b$-quark (or $\bar{b}$-quark) with momentum $p_\text{T}>5$ GeV/c and a radial separation from the reconstructed jet axis $\Delta R<0.3$. 
%\sout{In ATLAS measurements~[65-67], the jets are accepted in the rapidity range of $|y|<2.8$ for inclusive jet and $\gamma$+jet, $|y|<2.1$ for $b$-jet.}
In ATLAS measurements [65–67], inclusive jet and $\gamma$+jet
are reconstructed with $R=0.4$ and  accepted in the rapidity range of $|y|<2.8$,  while b-jet are reconstructed with $R=0.2$ and  accepted within $|y|<2.1$. Besides, for $\gamma$+jet event, $\gamma$ is required to have $p_\text{T}^\gamma>$ 50 GeV/$c$, and a cut $\Delta \phi_{\text{j}\gamma }>\pi/2$ is imposed to select the back-to-back $\gamma$+jet pairs. In our simulations, we implement correspondingly the same kinematic cuts adopted by experiments.

\begin{figure}[!t]
  \centering
  \vspace{15pt}
  \includegraphics[width=0.95\linewidth]{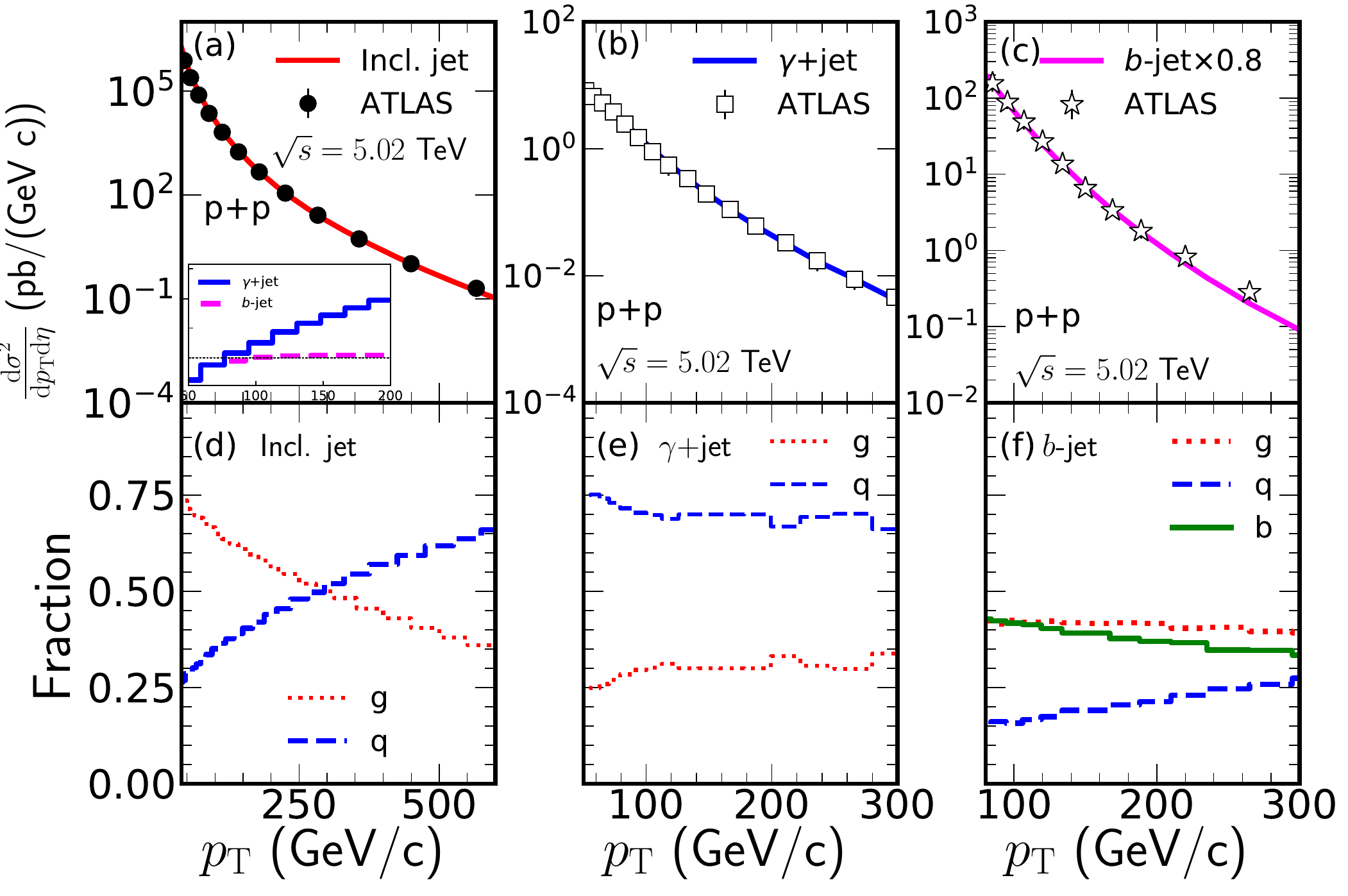}
  \caption{(Color online)  Up: Transverse momentum distributions of: (a) inclusive jet, (b) $\gamma$-tagged jet, and (c) $b$-jet simulated by MadGraph+Pythia8 (lines) and the comparison with experimental data (samples)~\cite{ATLAS:2018gwx,ATLAS:2022cim,ATLAS:2022fgb} in p+p collisions.  The
inset in (a) is the  ratio of $\gamma$-tagged jet (blue solid) and $b$-jet (red dashed) to inclusive jet cross section.
   Bottom: fraction of quark (Dashed blue line) and gluon (Solid red line) initiated jet of : (d) inclusive jet, (e) $\gamma$-tagged jet, and (f) $b$-jet  as a function of jet $p_\text{T}$ in p+p collisions.
   }\label{ref_pp_lbt}
\end{figure}

In the top panel of Fig.~\ref{ref_pp_lbt}, we plot the differential cross section of: (a) inclusive jet (denoted as
“Incl. jet” in the figure in the following), (b) $\gamma$+jet, and (c) $b$-jet  as a function of jet transverse momentum $p_\text{T}$ obtained from MadGraph+Pythia8 simulation at 5.02 TeV in p+p collisions. Through the comparison with experimental data~\cite{ATLAS:2018gwx,ATLAS:2022cim,ATLAS:2022fgb}, one can see clearly that the simulations give very well descriptions of all experimental data. Notice that the inset of Fig.~\ref{ref_pp_lbt}(a) is the scaled ratio of $\gamma$+jet (blue solid) and  $b$-jet (red dashed) cross section to that of inclusive jet. In Fig. \ref{ref_pp_lbt}(a-c), one can see that the inclusive jet spectrum is much steeper than $\gamma$+jet, while $b$-jet have similar slope as the inclusive jet, consistent with the results of Refs.~\cite{ATLAS:2022cim,ATLAS:2022fgb}.

In order to study the flavor dependence of jet energy attenuation in heavy ion collisions, we present the relevant contributions in terms of jet flavor, which is defined as the flavor of the hard parton that fragments into the final observed jet\footnote{If $\geq 2$ hard partons locate in the final observed jet, the flavor of a jet is defined as that of the hardest parton.}.
In the  bottom panel of Fig.~\ref{ref_pp_lbt}, we show the fraction from quark- and gluon- initiated jet in: (d) inclusive jet, (e) $\gamma$+jet, and (f) $b$-jet as a function of jet $p_\text{T}$. One can see that for inclusive jet, the contribution from gluon (quark) initiated jet dominates in low (large) $p_\text{T}$ region, and gradually decreases (increases) with increasing $p_\text{T}$. While for $\gamma$+jet, the quark initiated jet dominates ($\sim 80\%$) in the whole $p_\text{T}$ region. For $b$-jet, it can be generated either from the initial hard scattering or from the parton showers via gluon and quark splitting. In the first case, it is the $b$-quark that initiates the $b$-jet, the relevant contribution is shown by $b$-quark in Fig.~\ref{ref_pp_lbt}(f). In heavy-ion collisions, the medium modification to such $b$-jet has direct connection to the heavy quark energy loss~\cite{Zhang:2003wk,Djordjevic:2003qk,Armesto:2003jh,Dai:2022sjk}. On the other hand, the medium modification on the latter two cases (with gluon and quark splitting) would resemble that of a massive quark or gluon jets. As can be seen, $b$-jet from gluon initiated contributes about $40\%$ to the cross section in the whole $p_\text{T}$ region, while the light quark initiated contribution goes up with increasing $p_\text{T}$.

\begin{figure}
  \centering
  \includegraphics[width=0.5\textwidth]{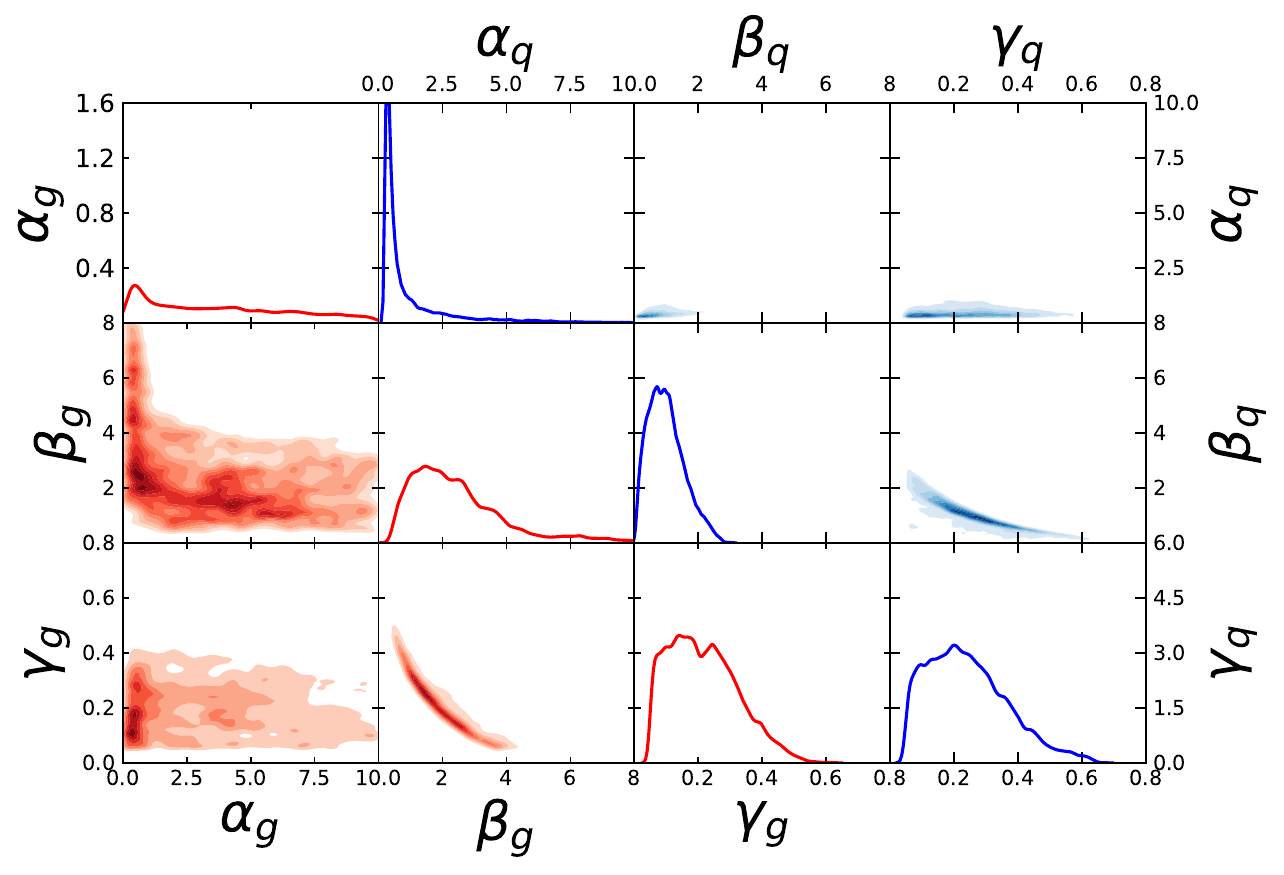}
  \caption{(Color online) Distributions of and the correlations between the Bayesian-extracted parameters for gluon (left) and quark (right) initiated jet energy loss via fitting to $R_\text{AA}$ of inclusive jet and $\gamma$-tagged jet  in central 0-10$\%$ Pb+Pb collisions at 5.02 TeV~\cite{ATLAS:2018gwx,ATLAS:2022cim}.  }\label{parameter}
\end{figure}

\subsection{Colour-charge dependence of $R_\text{AA}$}

In Fig.~\ref{parameter}, we present the distributions of the final-extracted parameters for gluon (left) and quark (right) initiated jet energy loss as well as their correlations, via Bayesian-fitting  to the ATLAS data~\cite{ATLAS:2018gwx,ATLAS:2022cim} on $R_\text{AA}$ of inclusive jets and $\gamma$-tagged jets in 0-10$\% $ Pb+Pb collisions at 5.02 TeV simultaneously.
%We do not take the effect of different original position of $\gamma$-tagged jet and inclusive jet  in the medium on the $R_\text{AA}$ into consideration at present.
As can be seen, $\beta_i$ and $\gamma_i$, which  reflect the average energy loss,  are strongly correlated and  well constrained for quark and gluon initiated jet.   The  mean value
as well as its standard deviation of those final extracted parameters for gluon and charm quark energy loss distribution are summarized in Table~\ref{table:nrqcd}.

\begin{figure}
  \centering
 \includegraphics[width=0.45\textwidth]{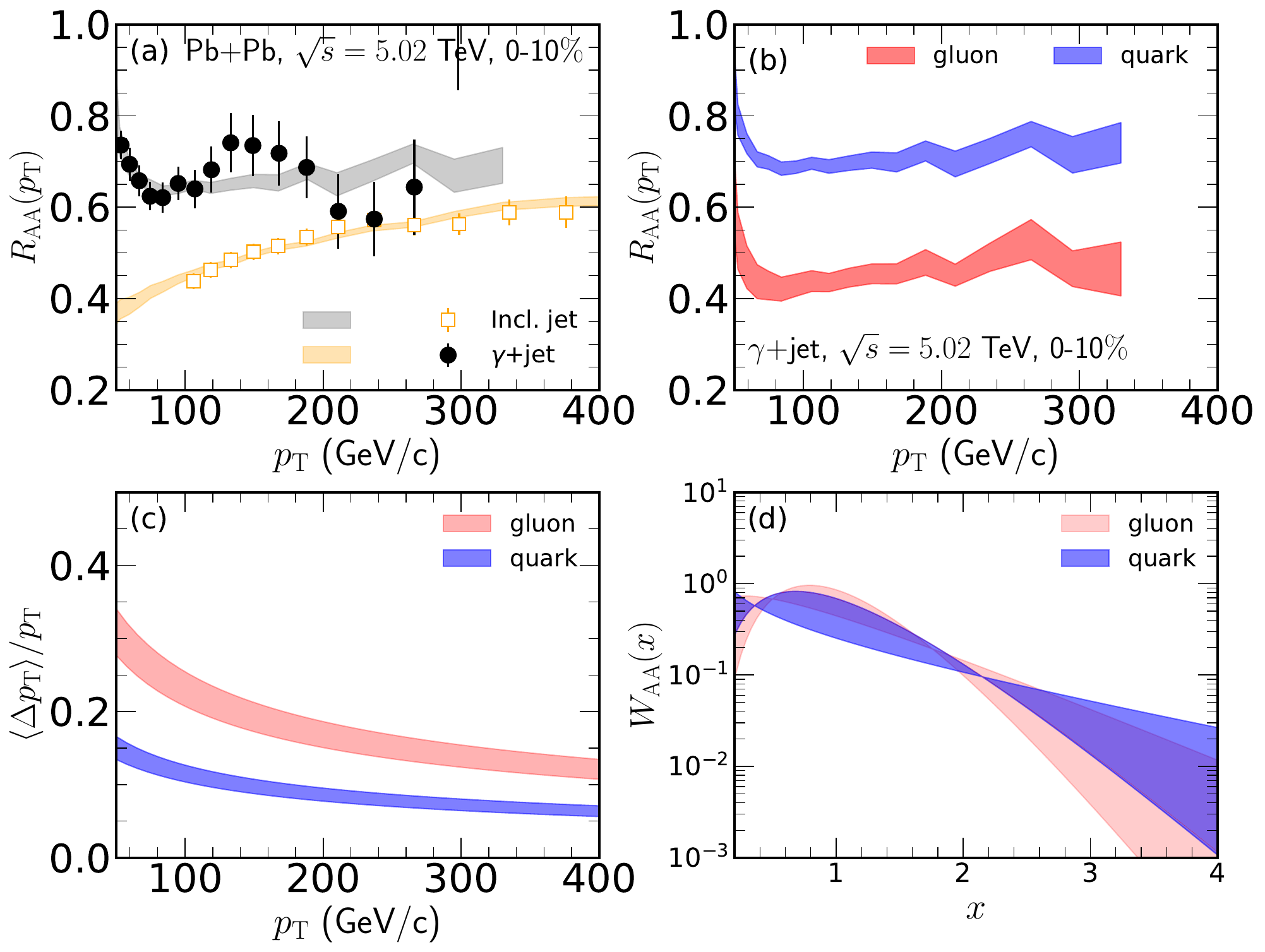}
  \caption{(Color online) (a) Data-driven Bayesian fitted  nuclear modification factor $R_\text{AA}$ of inclusive jet (orange) and $\gamma$-tagged jet (gray) and the comparison to experimental data at~\cite{ATLAS:2018gwx,ATLAS:2022cim}.  (b) Data-driven extracted  nuclear modification factor of quark (blue) and gluon (red) initiated $\gamma$+jet. (c) Fraction of jet average energy loss of light quark (blue) and gluon (red) initiated jet, (d) scaled energy loss distributions $W_\text{AA}^i(x)$ of  quark (blue) and gluon (red) initiated jet.  }\label{raa_ref}
\end{figure}

The final fitted nuclear modification factor $R_\text{AA}$ of inclusive jet  and $\gamma$-tagged jet as well as the comparison to experimental data~\cite{ATLAS:2018gwx,ATLAS:2022cim} in 0-10$\%$ centrality at 5.02 TeV are shown in Fig.~\ref{raa_ref}.(a), and data-driven extracted  nuclear modification factor of quark- and gluon- initiated $\gamma$+jet are shown in Fig.~\ref{raa_ref}.(b). The corresponding bands are results with one sigma deviation from the average fits of $R_\text{AA}$.
Considering the fact that the training process will minimize the Gaussian likelihood function between experimental data and the output for any given set of parameters, the final fitted results are almost close to the central value of data points. Moreover, considering the limited experimental data sets, our parametrization for the energy loss distribution shown in Eq. (\ref{Waa}) is limited to three parameters for each flavor, which could introduce correlations between different data bins. Therefore, the MCMC bands is more restricted than the uncertainty in the experimental data. We leave a more detailed analysis of the uncertainties for a future publication.
%The quark initiated jet  $R_\text{AA}$ and energy lose can be well constrained when only $\gamma$-tagged jet data  are taken into consideration, due to large fraction of quark jet, while gluon initiated jet $R_\text{AA}$ and energy lose are weakly constrained.  When experimental data of inclusive jet are included, the gluon jet  $R_\text{AA}$ and energy lose  can be further constrained.
Data-driven extracted average energy loss fraction $\langle \Delta p_\text{T}\rangle/ p_\text{T}$ and scaled energy loss distributions $W_\text{AA}(x)$ of quark and gluon initiated jet are also presented in Fig.~\ref{raa_ref}.(c) and Fig.~\ref{raa_ref}.(d).  %{\color{red}\sout{ The results from LBT model simulations are also presented in Fig.~\ref{raa_ref} for comparison. The agreement with LBT simulation validates the Bayesian analysis in extracting flavor dependent jet energy loss.}  }
%\sout{\color{red}Fig.~\ref{raa_ref} also show the LBT results which are in agreement with the Bayesian results. }
As can be seen, average energy loss of gluon  and quark jet is well constrained in $p_\text{T}<200$ GeV/$c$, but is weaker constrained at high $p_\text{T}$ due to large experimental data errors and the scarcity of  $\gamma$-tagged jet experimental data at such high $p_\text{T}$. The quark-initiated jets lose less fraction of its energy and shows a weaker dependence on the jet $p_\text{T}$ compared to gluon-initiated jets due to its color factor as expected.
Since jet showers also contain gluons even if they are initiated by a hard quark, the net energy loss of a gluon-tagged jet is always larger than that of a quark-tagged jet but the ratio is smaller than 9/4 from the naive leading order estimation~\cite{Wang:1998bha,Liu:2006sf,Chen:2008vha}.

%\section{Bayesian analysis}

\begin{figure}
  \centering
  \includegraphics[width=0.45\textwidth]{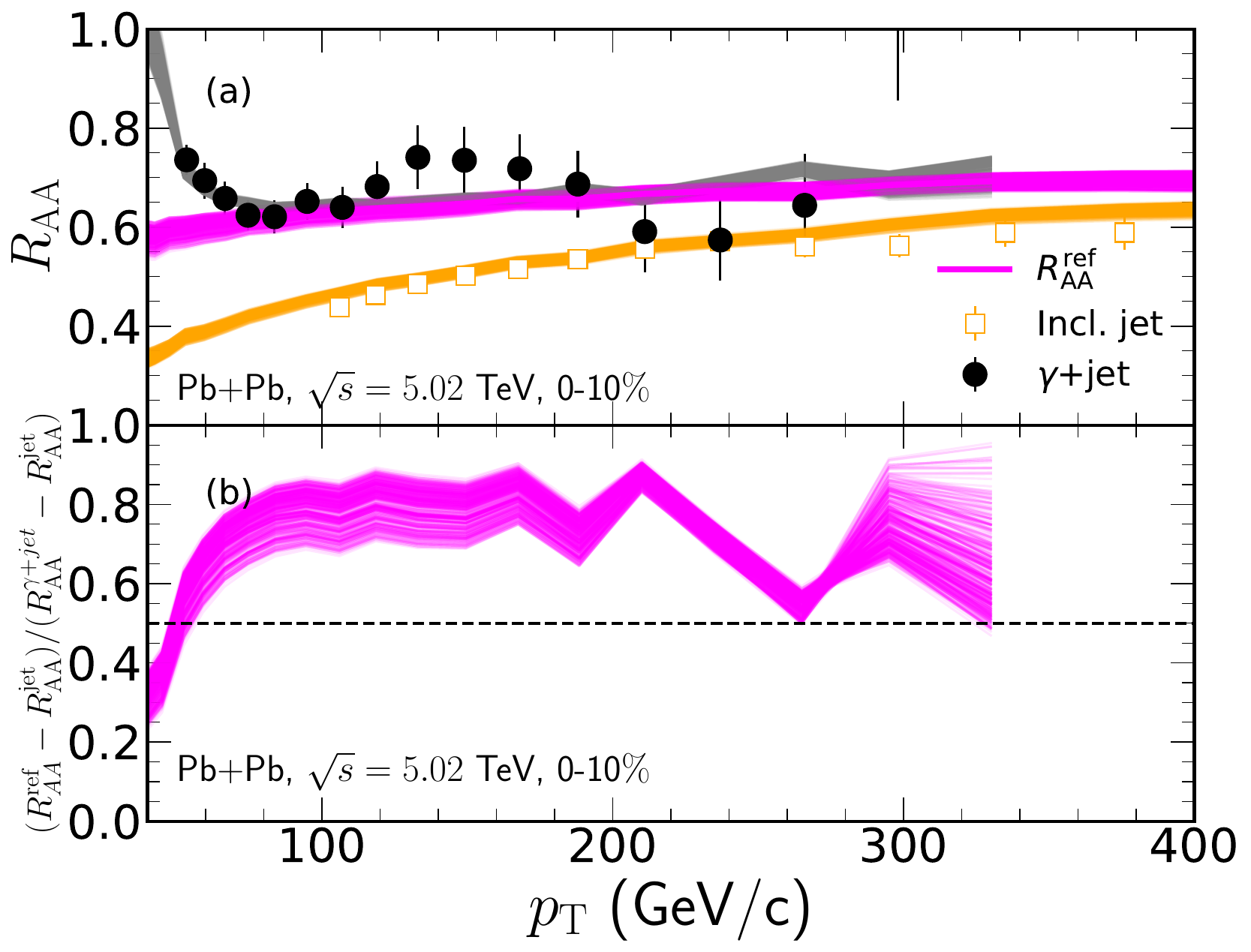}%RAA_c_\text{T}est   RAA_bjet_\text{T}est
  \caption{(Color online) (a) The reference $R_\text{AA}^{\text{ref}}$ (magenta) and the comparison with $R_\text{AA}$ of $\gamma$+jet (grey) and  inclusive jet (orange) jet in 0-10$\%$ centrality at 5.02 TeV, and the comparison with experimental data~\cite{ATLAS:2022cim}.  (b) The relative contribution fraction  from large quark fraction to the less suppression of $\gamma$+jet $R_\text{AA}$ compared to inclusive jet $R_\text{AA}$. }\label{Raa_gamma}
\end{figure}

Fig.~\ref{raa_ref}.(a) shows that 
 $\gamma$-tagged jet $R_\text{AA}$ is less suppressed compared to that for inclusive jet, which is a mix effect of the slope of initial spectra  and parton color-charge in p+p collisions.
To clarify the relative contributions from the color-charge effect and the initial parton spectra between $\gamma$-tagged jet and inclusive jet, we calculate an artificial reference $R_\text{AA}^{\text{ref}}$ following Eq.(\ref{EQ_raa}), by assuming the inclusive jet production has the same fraction of quark jet as $\gamma$+jet. This reference $R_\text{AA}^{\text{ref}}$ is shown by magenta lines in Fig.~\ref{Raa_gamma}.(a). 
The difference between $R_\text{AA}^{\text{ref}}$ and inclusive jet $R_\text{AA}$  (denoted as ``$R_\text{AA}^{\text{jet}}$")  should be attributed largely to the different color-charge effect between quark-medium and gluon-medium interactions, %  larger fraction of quark jet will lead to larger $R_\text{AA}$.
while the distinction between $R_\text{AA}^{\text{ref}}$ and $\gamma$+jet $R_\text{AA}$  (denoted as ``$R_\text{AA}^{\gamma\text{+jet}}$")  should be attributed mostly to the slope of reference spectra in p+p.

Fig.~\ref{Raa_gamma}.(b) shows the relative contribution fraction  from large quark fraction, evaluated as  $f^{\text{flavor}}=(R_\text{AA}^{\text{ref}}-R_\text{AA}^{\text{jet}})/(R_\text{AA}^{\gamma\text{+jet}}-R_\text{AA}^{\text{jet}})$, to the less suppression of $\gamma$+jet $R_\text{AA}$ compared to inclusive jet $R_\text{AA}$.  The increased quark jet fraction in inclusive jet production give the dominant contributions to the difference of  $R_\text{AA}$  between $\gamma$+jet and inclusive jet %$R_\text{AA}^{\gamma\text{+jet}}/ R_\text{AA}^{\text{jet}}$
at $p_\text{T}>60$ GeV/$c$. Then $1-f^{\text{flavor}}$ characterized approximately  the relative contribution from the slope of reference spectra, which plays a dominated role in the suppression at low $p_\text{T}$.
Besides, the distinction between $\gamma$+jet $R_\text{AA}$ and inclusive jet $R_\text{AA}$ will diminish with increasing $p_\text{T}$, because quark-initiated jets contribute a lion's share to the yields of both $\gamma$+jet and the inclusive jet at very large $p_\text{T}$, which can be verified with the upcoming high precision measurements at the LHC.
%The ratio of extracted nuclear modification factor $R_\text{AA}$ of $\gamma$+jet to inclusive jet as well as $R_\text{AA}^r/R_\text{AA}^{inclusive \ jet}$ in 0-10$\%$ centrality are also shown in Fig.~\ref{Raa_gamma}.(b). The artificial increased quark jet fraction in inclusive jet production leads to more than $80\%$ contributions to the ratio of $R_\text{AA}^{\gamma\text{+jet}}/ R_\text{AA}^{inclusive\  jet}$ in $80<p_\text{T}^{jet}<200$ GeV region, while the slope of reference spectra dominate in low $p_\text{T}$ region.\is{(don't quite understand this statement)} Besides, the difference between $\gamma$+jet $R_\text{AA}$ and inclusive jet $R_\text{AA}$ will decrease with increasing $p_\text{T}$, because the quark/ gluon fraction tend to be similar in large $p_\text{T}$ region, which can be verified with the upcoming high precision measurements at LHC.% \is{(how about the slop in large pt?)}.

\subsection{Centrality dependence of $R_\text{AA}$}
\begin{figure}
  \centering
  \includegraphics[width=0.5\textwidth]{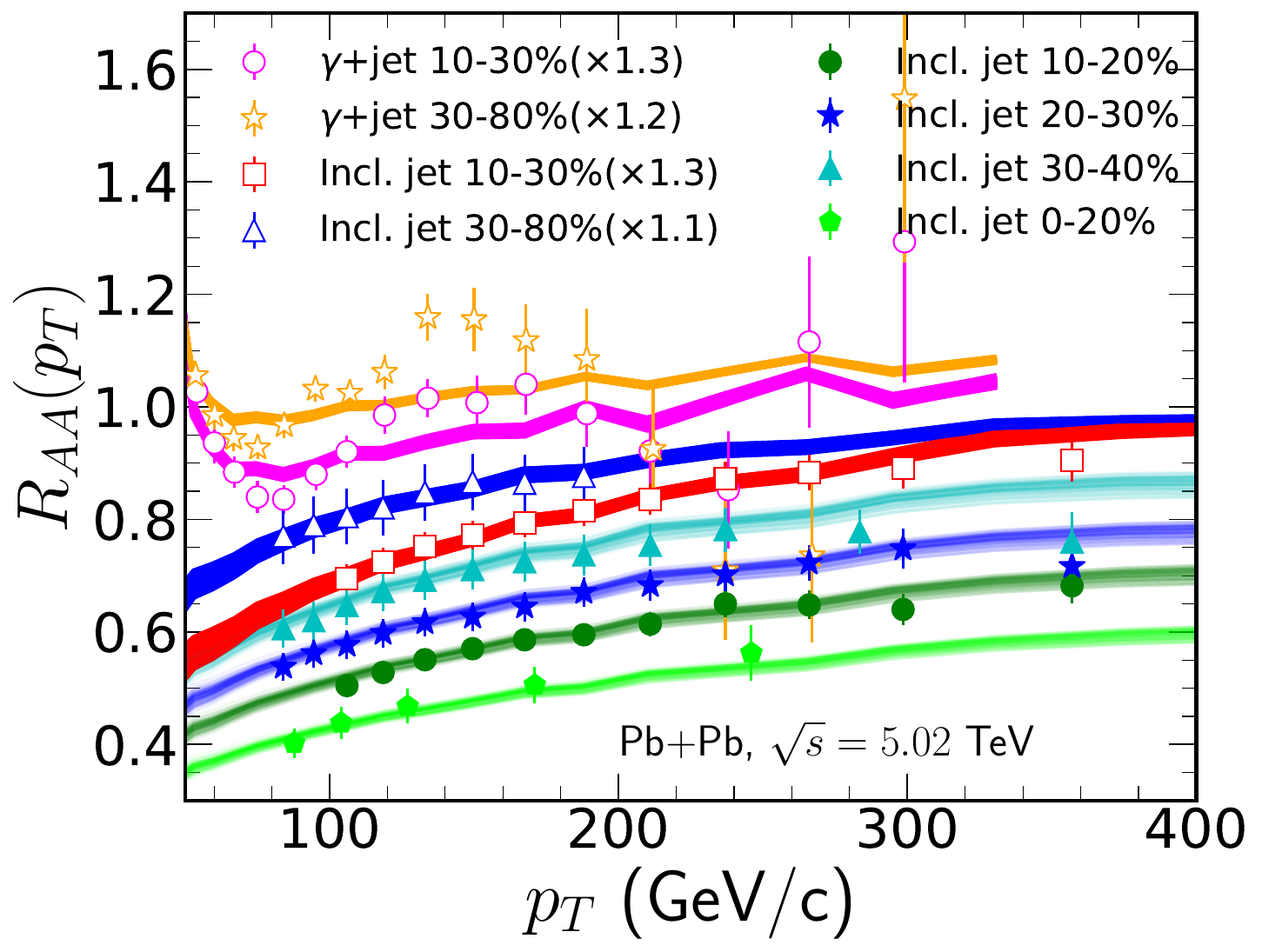}
  \caption{(Color online) Data-driven fitted  nuclear modification factor $R_\text{AA}$ of the inclusive jet 
  $R_\text{AA}$~\cite{ATLAS:2018gwx} and  $\gamma$+jet $R_\text{AA}$~\cite{ATLAS:2022cim} in 10-30$\%$, 30-80$\%$ centrality bins  and predictions of inclusive jet $R_\text{AA}$ in 10-20$\%$, 20-30$\%$, 30-40$\%$ and 0-20$\%$   centrality bins as well as  the comparison with experimental data~\cite{ATLAS:2018gwx}.
 }\label{Raa_c_fit}
\end{figure}

\begin{footnotesize}
\begin{table}[!t]
\caption{Parameters [$\alpha_i$, $\gamma_i$, $\beta_i$] of quark and gluon jet energy loss distribution from Bayesian fits to experimental data~\cite{ATLAS:2018gwx,ATLAS:2022cim} on inclusive jet and $\gamma$+jet suppressions at 5.02 TeV. }
\label{table:nrqcd}
\begin{center}
  \begin{tabular}{l|c c c c}
  \hline
    &  &  {$\alpha_i$ } &   {$\beta_i$} & {$\gamma_i$}  \\
   \hline
   \multirow{2}*{0-10$\%$} & gluon & 4.36$\pm$2.07    & 1.78$\pm$0.38   &    0.25$\pm$0.03 \\
                           ~&quark & 0.5$\pm$0.07    & 0.39$\pm$0.17   &    0.32$\pm$0.13 \\ \hline

    \multirow{2}*{10-30$\%$} & gluon & 2.17$\pm$0.94    & 1.47$\pm$0.44   &    0.25$\pm$0.04 \\
                             ~&quark & 5.81$\pm$1.8    & 1.27$\pm$0.12   &    0.09$\pm$0.02 \\ \hline

    \multirow{2}*{30-80$\%$} & gluon & 4.78$\pm$1.87    & 1.16$\pm$0.17   &    0.11$\pm$0.03 \\
                             ~&quark & 6.4$\pm$2.63    & 0.7$\pm$0.05   &    0.09$\pm$0.01 \\ \hline
  % Chao 1 & 1.16$\pm$0.2 & 8.9$\pm$0.98 & 0.30$\pm$0.12  & 1.26$\pm$0.47 \\ \hline
  % Chao 2 & 1.16$\pm$0.2 & 11 &  0  & 0 \\ \hline
  % Gong & 1.16$\pm$0.2 & 9.7$\pm$0.9 &  -0.46$\pm$0.13 & -2.14$\pm$0.56 \\ \hline
  \end{tabular}
\end{center}
\end{table}
  \end{footnotesize}

Moreover, we extract the centrality dependent  quark and gluon jet energy loss distributions before exploring parton-mass effect on jet quenching  motivated by two reasons.
First, $\gamma$-tagged jet $R_\text{AA}$~\cite{ATLAS:2022cim} shows a weaker dependence on centrality compared to inclusive jet~\cite{ATLAS:2018gwx}, indicating that gluon-initiated jets may show a distinct centrality dependence with quark-initiated jets.
%Besides, we can not well constrain $b$-quark jet energy loss, light-quark and gluon initiated jet energy loss, when  only b\textrm{-}jet $R_\text{AA}$ and inclusive jet $R_\text{AA}$  are  included in the fitting due to their subdominant contributions  and the limited data points with large uncertainties.
Second, the experimental data of $\gamma$+jet $R_\text{AA}$~\cite{ATLAS:2022cim}, inclusive jet $R_\text{AA}$~\cite{ATLAS:2018gwx}  and $b$-jet  $R_\text{AA}$~\cite{ATLAS:2022fgb} are in different centrality bins.
We need centrality dependent quark and gluon jet energy loss distributions to fit $\gamma$+jet $R_\text{AA}$, inclusive jet $R_\text{AA}$  and $b$-jet  $R_\text{AA}$ simultaneously.

As a matter of fact, there are no experimental data of inclusive jet $R_\text{AA}$ and $\gamma$+jet $R_\text{AA}$ in the same centrality class except in central 0-10$\%$ centrality. For inclusive jet measurements, the existing measurements are provided for centrality bins 0-10$\%$, 10-20$\%$, 20-30$\%$,
30-40$\%$, 40-50$\%$,50-60$\%$, 60-70$\%$, 70-80$\%$~\cite{ATLAS:2018gwx}, while for $\gamma$+jet $R_\text{AA}$, it is  limited to 0-10$\%$, 10-30$\%$, 30-80$\%$~\cite{ATLAS:2022cim}. In order to take full advantage of the existing measurements for inclusive jet $R_\text{AA}$ in different centrality bins, we generate the inclusive jet $R_\text{AA}$ as well as the corresponding errors in 10-30$\%$ and 30-80$\%$ centrality bins according to $R_\text{AA}^C=\sum_{c\in C} P^{c} R_\text{AA}^c$, where $P^c =N^c_\text{bin}/\sum_c N^c_\text{bin}$ is the probability of finding jet events in a given centrality bin following Ref.~\cite{Xing:2019xae}. With such an extension, we can perform a simultaneous fit for both inclusive jet $R_\text{AA}$ and $\gamma$+jet $R_\text{AA}$ in 10-30$\%$ and 30-80$\%$ centralities. In Fig.~\ref{Raa_c_fit}, we present the data-driven fitted nuclear modification factor $R_\text{AA}$ of inclusive jet~\cite{ATLAS:2018gwx} and $\gamma$+jet ~\cite{ATLAS:2022cim} in 10-30$\%$ and 30-80$\%$ centralities and the comparison with experimental data at 5.02 TeV. %\iss{Similarly, those procedures are also applied  to inclusive jet $R_\text{AA}$ and $\gamma$+jet $R_\text{AA}$ in 30-80$\%$ centrality and the final fitted results are also shown in  Fig.~\ref{Raa_c_fit}.}
All final spectra based on Eq.~(\ref{EQ_raa}) and Eq.~(\ref{sigma_AQ}) are in nice agreement with the experimental data. The corresponding mean value
as well as its standard deviation of those final extracted parameters for gluon and light quark energy loss distribution are summarized in Table~\ref{table:nrqcd}.

\begin{figure}[t]
  \centering
  \includegraphics[width=0.5\textwidth]{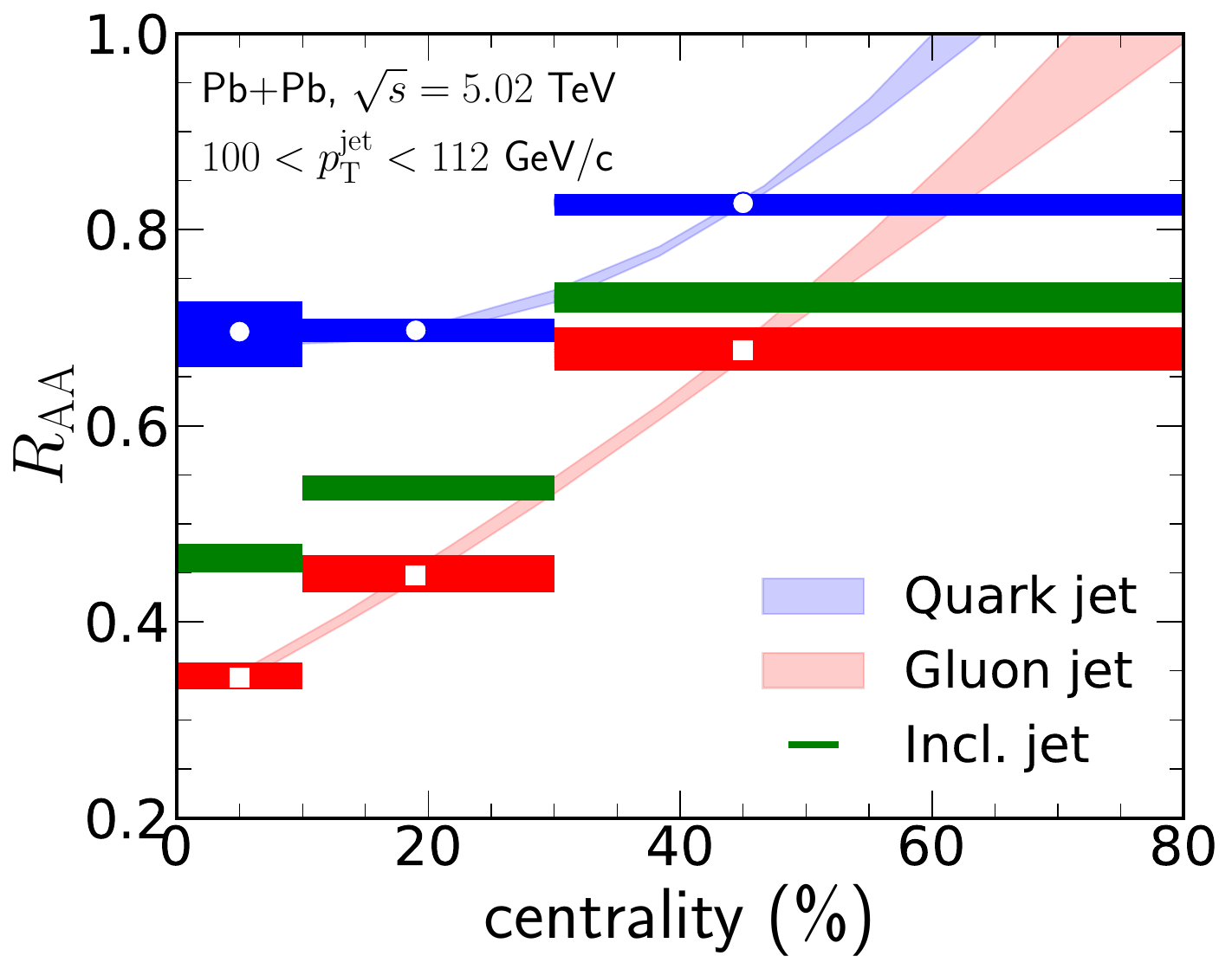} %\includegraphics[width=0.5\textwidth]{./RAA_c_pre.pdf}
  \caption{(Color online) The centrality dependence of final fitted  gluon jet (red), quark jet (blue) and inclusive jet (green) $R_\text{AA}$ in Pb+Pb collisions at 5.02 TeV.  }\label{Raa_c}
\end{figure}

Meanwhile, we obtain $R_\text{AA}$ for quark-initiated jets and gluon-initiated jets  in 10-30$\%$, 30-80$\%$ centrality. Combined with the flavor dependent $R_\text{AA}$ in 0-10$\%$ as extracted in the previous section (Fig.~\ref{raa_ref}.(b)), we obtain the centrality dependence of final
fitted gluon-initiated jet, quark-initiated jet and inclusive jet $R_\text{AA}$.  In Fig.~\ref{Raa_c}, we show the centrality dependence of final
fitted gluon jet (red), quark jet (blue) and inclusive jet (green)
$R_\text{AA}$ in Pb+Pb collisions  in the region $100<p_\text{T}<112 $ GeV/$c$ by step lines. One finds that the quark-initiated jet has weaker dependence on the centrality than that for gluon-initiated jet. %The open symbols are the centrality corresponding to the impact parameters in the three different centrality bins respectively. \is{(The $R_\text{AA}^{\text{jet}}$ in 30-80$\%$ is similar equal to that for 40-50$\%$.   )}.

Next, we can fit the centrality dependent $R_\text{AA}$ of quark- and gluon- initiated jet via a simple parametrization $h^i(C)=a_i C^2+b_i C+c_i$, with $C$ stands for the centrality. The best fit curves of $h^i(C)$ are shown in Fig.~\ref{Raa_c} by blue and red band, and the corresponding best-fit parameter values are presented in Table.~\ref{table:parameters}.
%\iss{The fitted $f^i(c) $ is greater than one and unreasonable in peripheral collisions, which need to be further constrained  by  more centrality dependent $\gamma$-tagged jet  experimental data.}
Notice that the extrapolation to peripheral collisions ($>80\%$) is greater than one and can not be trusted, a reasonable identification of jet energy loss distribution for peripheral collisions would require a corresponding extension of experimental measurements.
If we ignore the $p_\text{T}$ dependence of $h^i(C)$, $R_\text{AA}^{i,C}$ for any centrality $C$ can be simply obtained by $R_\text{AA}^{i,C}=h^i(C)*R_\text{AA}^{i,rc}/h^i(rc)$, where $rc$ stands for reference centrality. Based on Eq.~(\ref{EQ_raa}) and the above extracted centrality dependent quark and gluon jet $h^i(C)$, the predication of inclusive jet $R_\text{AA}$ in 0-20$\%$, 10-20$\%$, 20-30$\%$, 30-40$\%$ are presented in Fig.~\ref{Raa_c_fit}. One can see that our extracted centrality dependence of quark and gluon jet energy loss distributions can describe the experimental data  $R_\text{AA}$~\cite{ATLAS:2018gwx} very well.
%\is{there are pt-dependence in those curves, need to explain the pt-dependence of $f^i(C)$.}

\begin{table}[!t]
  \caption{The best-fitted Parameters [$a_i$, $b_i$, $c_i$] for centrality dependent quark and gluon jet energy loss distributions. }
\begin{center}
 % \begin{tabular}{l|c|c|c|c}
  \begin{tabular}{p{1.0cm}<{\centering}|p{2.5cm}<{\centering}|p{2.2cm}<{\centering}|p{2.cm}<{\centering}}

   \hline
            & $a_i \  (\times 10^{-5})$          &    $b_i \ (\times 10^{-3}) $        &  $c_i$    \\ \hline
    Quark & 12.39$\pm$2.83    &    -2.95$\pm$1.74 &   0.7$\pm$0.021   \\ \hline
   Gluon &3.36$\pm$2.45   &  6.65$\pm$1.20   & 0.309$\pm$0.00879 \\  \hline
  \end{tabular}

  \label{table:parameters}
\end{center}
\vspace{-10pt}
\end{table}

\begin{table}[!t]
  \caption{Parameters [$\alpha_i$, $\gamma_i$, $\beta_i$] for gluon, quark and $b$-quark  energy loss in 0-20$\%$  centrality from  fitting to  $\gamma$+jet $R_\text{AA}$~\cite{ATLAS:2022cim}, inclusive jet $R_\text{AA}$~\cite{ATLAS:2018gwx}  and $b$-jet  $R_\text{AA}$~\cite{ATLAS:2022fgb} at $\sqrt{s}=5.02$ TeV simultaneously. }
\begin{center}
 % \begin{tabular}{l|c|c|c|c}
  \begin{tabular}{p{1.5cm}<{\centering}|p{2.cm}<{\centering}|p{2.cm}<{\centering}|p{2.cm}<{\centering}}
  \hline
  \multicolumn{4}{c} {(0-20$\%$) 5.02 TeV }   \\
   \hline
            & $\alpha_i$          &    $\beta_i $        &  $\gamma_i$    \\ \hline
    Gluon & 4.60$\pm$2.96   &    2.18$\pm$1.12 & 0.21$\pm$0.12  \\ \hline
    Quark & 4.12$\pm$2.71   &  0.86$\pm$0.38   & 0.24$\pm$0.11  \\ \hline
    $b$-quark &5.32$\pm$2.84   &  0.80$\pm$0.54   & 0.2$\pm$0.17  \\ \hline
  \end{tabular}

  \label{table:parameters_b}
\end{center}
\vspace{-10pt}
\end{table}

\subsection{Parton-mass  dependence of $R_\text{AA}$}
Finally, with the extracted centrality dependent quark and gluon energy loss distributions, we also extract $b$-jet energy loss in the same framework based on  Eqs.~(\ref{EQ_raa}) and (\ref{sigma_AQ}) through fitting to the experimental data of $b$-jet $R_\text{AA}$~\cite{ATLAS:2022fgb}, inclusive jet $R_\text{AA} $ in  0-20$\%$ centrality~\cite{ATLAS:2018gwx} and $\gamma$-tagged jet $R_\text{AA}$ in  0-10$\%$~\cite{ATLAS:2022cim} simultaneously.
 %{\color{blue} $b$-jets $R_\text{AA}$, inclusive jet $R_\text{AA} $ and $\gamma$-tagged jet $R_\text{AA}$ are in different centrality, so we need  centrality dependence of quark and gluon energy lose distributions to fit those $R_\text{AA}$ simultaneously. For example, $\gamma$-tagged jet $R_\text{AA}$ in 0-20$\%$ centrality, which is generated during the fitting,  is changed to $R_\text{AA}$ in 10-30$\%$ to compare with experimental data directly.}.
Considering the recent CMS measurements~\cite{CMS:2016uxf,CMS:2021vui}, as well as an earlier ATLAS measurement~\cite{ATLAS:2012tjt}, where the ratio of inclusive jet $R_\text{AA}$ with jet cone R=0.4 to R=0.2 show no deviation from one at large $p_\text{T}$, and  the limited b-jet data points,  we ignore the jet cone dependence at present and  mainly focus on a qualitative investigation on the parton mass/flavor effects on the b-jet $R_\text{AA}$.

\begin{figure}
  \centering
  \includegraphics[width=0.5\textwidth]{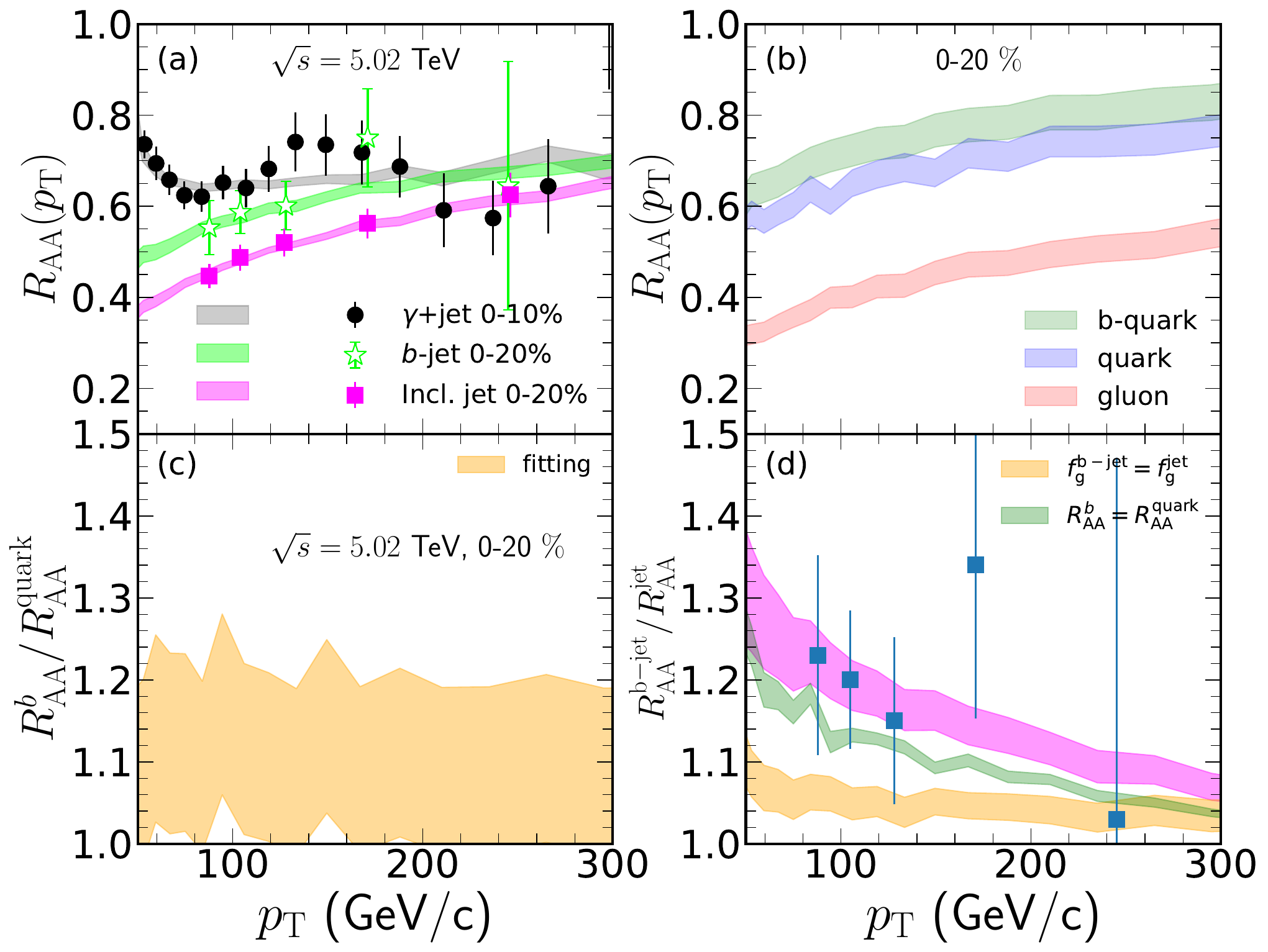}   % RAA_bjet_\text{T}est   RAA_bjet_cpmparison
  \caption{(Color online) (a): final fitted nuclear modification
factor $R_\text{AA}$ of  $b$-jets (lime green), inclusive jet (magenta)  and $\gamma$-tagged jet (gray) and the comparison with experimental data~\cite{ATLAS:2018gwx,ATLAS:2022cim,ATLAS:2022fgb}. (b): the data-driven extracted $R_\text{AA}$ of gluon (red), light quark (blue), and $b$-quark (green) initiated jets.
  (c): The ratio of $R_\text{AA}^{b}/R_\text{AA}^\text{quark}$ from  global analysis.
  (d):  the ratio of $R_\text{AA}^{b\text{-jet}}/R_\text{AA}^{\text{jet}}$  extracted  from  global analysis and the comparison to the experimental measurements.  The quark mass effect (yellow) and less gluon fraction effect (green)  to the ratio  of $R_\text{AA}$ are also presented.  %{\color{red}\sout{The solid, dotted and dashed lines are from LBT simulations.}} 
  }\label{Raa_bjet}
\end{figure}

The final fitted  nuclear modification factor $R_\text{AA}$ of $b$-jet (lime green band), inclusive jet (magenta band)  and $\gamma$-tagged jet (gray band) as well as the comparison with experimental data~\cite{ATLAS:2018gwx,ATLAS:2022cim,ATLAS:2022fgb} are shown in Fig.~\ref{Raa_bjet}(a). The corresponding bands are results with one sigma deviation from the average fits of $R_\text{AA}$.
Meanwhile, Fig.~\ref{Raa_bjet}(b) shows the extracted nuclear modification factor $R_\text{AA}$ for $b$-quark initiated (green, denoted as ``$R_\text{AA}^{b}$"), light-quark (denoted as ``$R_\text{AA}^{\text{quark}}$") and gluon (denoted as ``$R_\text{AA}^{\text{gluon}}$") initiated $b$-jet in 0-20$\%$ centrality, with the corresponding  parameters for gluon, quark and $b$-quark energy loss distribution summarized in Table.\ref{table:parameters_b}.  
The final extracted light-quark and gluon initiated jet energy loss distributions are consistent with our previous results in the same centrality, while $b$-quark initiated jets is less suppressed compared to light-quark initiated jets due to its large mass in low $p_\text{T}$ region.
 Our result shows
a clear flavor hierarchy of jet energy loss at high energies, $\langle \Delta E_g \rangle > \langle\Delta E_q\rangle > \langle\Delta E_b\rangle$ inside a hot
nuclear matter, consistent with perturbative QCD expectation. 
%\sout{To verify the rationality of the extracted flavor dependence of jet quenching, we compare the results to LBT simulations shown in lines in Fig.~\ref{Raa_bjet}. The agreement between Bayesian analysis and LBT validates the expected flavor hierarchy of jet quenching, i.e. $\langle \Delta E_g \rangle > \langle\Delta E_q\rangle > \langle\Delta E_b\rangle$.}

To explore the underlying $b$-jet suppression mechanism in heavy-ion collisions, we also present in Fig.~\ref{Raa_bjet}(c) the ratio of  $b$-quark initiated jets $R_\text{AA}$ to light quark initialed 
jet $R_\text{AA}$ as $R_\text{AA}^{b}/R_\text{AA}^{\text{quark}}$, and also in Fig.~\ref{Raa_bjet}(d) the ratio of $b$-jet $R_\text{AA}$ (denoted as ``$R_\text{AA}^{b\text{-jet}}$") to inclusive jet $R_\text{AA}$ $R_\text{AA}^{b\text{-jet}}/R_\text{AA}^{\text{jet}}$  extracted  from  global analysis  and the comparison  to the experimental measurements~\cite{ATLAS:2022fgb}. Our numerical results can describe the experimental data within large uncertainties~\cite{ATLAS:2022fgb}. Those ratios are greater than unity and  go down with increasing $p_\text{T}$, indicating that the parton mass effect is reduced with increasing  $p_\text{T}$~\cite{Huang:2013vaa}.
However, the mass effect for $b$-jet could persist to large $p_\text{T}$, even at $p_\text{T}\sim300$ GeV/$c$, and is consistent with the current data and a model based on strong coupling (via the
AdS/CFT correspondence)~\cite{Horowitz:2007su}, in contrast to Ref.~\cite{Huang:2013vaa,Xing:2019xae}  in which mass effects are
expected to be small at  $p_\text{T}>70 $ GeV/$c$. Those disagreements  may be explained  by two reasons. For one thing, due to the subdominant contributions and the limited $b$-jet $R_\text{AA}$ data points with large uncertainties, especially at large $p_\text{T}$, which have weak constraints on $b$-quark initialed jet, $b$-quark initiated jet energy loss distributions is weakly constrained at present.  For another, this may be attributed to the mixture of mass effect and color effect, as we may show below.

%In fact,  \is{(why we don't see the consistency between light quark and b-quark in the large pt region? one should expect the vanish of mass effect in large pt region such as 300 GeV)}

%This may be attributed to the jet substructures.  The splitting function of $b$-quark peaks at $z\sim1$ in PYTHIA. Therefore, b-quark initiated jets are harder and lazier than light-quark/gluon initiated jets. Jets with more complex constituents tend to lose more energy compared to those jets with simpler substructures. Thus  $b$-initiated jets will lose less energy compared to quark/gluon jets due to the substructures, which can be verified though the comparison of jet fragmentation function $z=p_\text{T}^{h}/p_\text{T}^j$ of light and $b/c$-jets. \is{this argument is not clear enough.}
Notice that b-jet spectra~\cite{ATLAS:2022fgb} have similar slope as the inclusive jet as shown in the inset of Fig.  1(a),  so the 
 parton mass and  color effects may give the dominated contributions  to the  difference between inclusive jet $R_\text{AA}$ and $b$-jet $R_\text{AA}$.
 For further demonstration of $b$-quark mass effect on the suppression of $b$-jet, we show in Fig.~\ref{Raa_bjet}(d) (yellow band) the ratio of $b$-jet $R_\text{AA}$ to inclusive jet $R_\text{AA}$, assuming $b$-jet has the same fraction of gluon initiated jet as inclusive jet (denoted as ``$f^{b\text{-jet}}=f^{\text{jet}}$").  The difference between this ratio and  $R_\text{AA}^{b\text{-jet}}/R_\text{AA}^{\text{jet}}$
should be attributed to the $b$-quark mass effect.
 One can see that the deviation between $b$-jet and the inclusive jet is moderately reduced with increasing $p_\text{T}$. %Despite $b$-initiated jets is less suppressed compared to light-quark initiated jets,
 The mass effect roughly give considerable contributions to the ratio of $R_\text{AA}^{b\text{-jet}}/R_\text{AA}^{\text{jet}}$ and are expected to be small at $p_\text{T} \sim 300 $ GeV/$c$.

To further illustrate the color-charge effect on the suppression of $b$-jet,  we also calculated the  ratio of $b$-jet $R_\text{AA}$ to inclusive jet $R_\text{AA}$, assuming $b$-quark jet lose the same fraction of energy as light-quark  initiated jet (denoted as ``$R_\text{AA}^{b}=R_\text{AA}^{\text{quark}}$"), as shown by green band in Fig.~\ref{Raa_bjet}(d). The difference between this ratio and  $R_\text{AA}^{b\text{-jet}}/R_\text{AA}^{\text{jet}}$
should be attributed to the different gluon and quark fraction.
 As can be seen, those ratio is significantly enhanced and also show a downward tendency with  increasing $p_\text{T}$, indicating that, less gluon initiated jet contribution also lead to the less suppression of $b$-jet compared to inclusive jet in heavy-ion collisions, especially in low $p_\text{T}$ region.  Therefore, we can see that the color charge effect have greater impacts to the ratio $R_\text{AA}^{b\text{-jet}}/R_\text{AA}^{\text{jet}}$ than parton mass effect in heavy-ion collisions.
 Furthermore, the contribution from gluon initialed jet to inclusive jet production is greater than that to $b$-jet in the $p_\text{T}<300$ GeV/$c$ region  as shown in Fig.~\ref{ref_pp_lbt}. Thus $b$-jet $R_\text{AA}$ is moderately larger than inclusive jet $R_\text{AA}$.

\section{Summary}
\label{summary}
 We have carried out a systematic  investigation of parton color-charge  and parton mass dependence of nuclear modification factor by a systematic study  of  the medium modifications on  three full jet observables: the inclusive  jet, $\gamma$+jet, and $b$-jet, in Pb+Pb collisions relative to that in p+p at the LHC.  Our  results  from MadGraph+PYTHIA  give very nice descriptions of the experimental data for these three jet observables  in p+p.  Then a Bayesian data-driven method is applied to extract the model-independent but
flavor-dependent jet energy loss distributions. Fitting to those experimental data simultaneously, we present the first quantitative extraction of  gluon, light quark and $b$-quark initiated jet energy loss distributions in  heavy-ion collisions.  It is seen that the energy loss of quark-initiated jets shows a weaker centrality dependence and weaker $p_\text{T}$ dependence compared to that of gluon-initiated jets. Our result shows
a clear flavor hierarchy of jet energy loss at high energies, $\langle \Delta E_g \rangle > \langle\Delta E_q\rangle > \langle\Delta E_b\rangle$ inside a hot
nuclear matter, consistent with perturbative QCD expectation. % {\color{red}\sout{and the LBT simulations}}.

Furthermore, we analysed the relative contributions from the
slope of initial spectra, parton color-charge as well as parton
mass dependent jet energy attenuation to the $\gamma$/b-jet suppression in heavy-ion collisions.
 We find that large quark-initiated jet fraction  underlies $\gamma$+jet suppression at large $p_\text{T}$, while the flat spectra give the dominate contribution to $\gamma$+jet suppression at low $p_\text{T}$. We demonstrate that the color charge effect have greater impacts to the ratio $R_\text{AA}^{b\text{-jet}}/R_\text{AA}^{\text{jet}}$ than parton mass effect,  which decrease moderately  at $p_\text{T} \sim 300 $ GeV/$c$.
 %Except for mass effect, smaller fraction of gluon contributions lead to smaller suppression of $b$-jet compared to inclusive jet in low $p_\text{T}$ jet region.
Such a systematic extraction of
jet energy loss distributions can help constrain model
uncertainties  and pave the way to precise predictions of
the properties of the hot QCD medium created in relativistic heavy-ion collisions\footnote{When finalizing this paper, the authors notice a very recent parallel study of extracting the flavor dependence of parton energy loss~\cite{Xing:2023ciw}, but from the nuclear modifications of various hadron species instead of jet observables presented in our work.}.
%%%%%%%%%%%%%???%%%%%%%%%%%%%%%%

Several caveats should be mentioned. First, due to the limited data, we ignore the jet cone dependence~\cite{CMS:2021vui,ATLAS:2012tjt} at present and mainly focus on a qualitative investigation on the effects from the initial spectrum and parton mass/flavor on the jet $R_\text{AA}$. For a more strict study, we should use measurements with the same R in a global analysis. With
the upcoming high precision measurements of b-jet $R_\text{AA}$ at the LHC, one can quantitatively  analyse those mass/ flavor dependent jet quenching. 
%Second, the impact of the isospin and nPDF effects for the three processes are ignored in our analysis.  Even though the isospin effect for inclusive jets is negligible, we should take the impact of the isospin and nPDF effects for the other two processes into consideration. 
Second, the MCMC bands is more restricted than the uncertainty in the experimental data. A more detailed treatment in the Bayesian analysis of the uncertainties is needed in the future. Finally,  the Bayesian analysis here uses specific functional form for the parameterzation, thus introducing long-range correlations in the parameter space which may potentially bias the extracted parameters. A possible solution to tackle such issue is to use information field based approach as presented in Ref.~\cite{Xie:2022ght}.

{\bf Acknowledgments:} This research is supported by National Natural Science Foundation of China with Project Nos. 12035007, 12022512, 12147131, Guangdong Major Project of Basic and Applied Basic Research No. 2020B0301030008. S.Z. is further supported by the MOE Key Laboratory of Quark and Lepton Physics (CCNU) under Project
No. QLPL2021P01.

\end{document}